# Choosing a suitable acquisition function for batch Bayesian optimization: comparison of serial and Monte Carlo approaches†

Imon Mia,[a] Mark Lee,[b] Weijie Xu,[a] William Vandenberghe[a] and Julia W. P. Hsu *[a]

Batch Bayesian optimization is widely used for optimizing expensive experimental processes when several samples can be tested together to save time or cost. A central decision in designing a Bayesian optimization campaign to guide experiments is the choice of a batch acquisition function when little or nothing is known about the landscape of the "black box" function to be optimized. To inform this decision, we first compare the performance of serial and Monte Carlo batch acquisition functions on two mathematical functions that serve as proxies for typical materials synthesis and processing experiments. The two functions, both in six dimensions, are the Ackley function, which epitomizes a "needle-in-haystack" search, and the Hartmann function, which exemplifies a "false optimum" problem. Our study evaluates the serial upper confidence bound with local penalization (UCB/LP) batch acquisition policy against Monte Carlo-based parallel approaches: $q$-log expected improvement ($q$logEI) and $q$-upper confidence bound ($q$UCB), where $q$ is the batch size. Tests on Ackley and Hartmann show that UCB/LP and $q$UCB perform well in noiseless conditions, both outperforming $q$logEI. For the Hartmann function with noise, all Monte Carlo functions achieve faster convergence with less sensitivity to initial conditions compared to UCB/LP. We then confirm the findings on an empirical regression model built from experimental data in maximizing power conversion efficiency of flexible perovskite solar cells. Our results suggest that when empirically optimizing a "black-box" function in ≤six dimensions with no prior knowledge of the landscape or noise characteristics, $q$UCB is best suited as the default to maximize confidence in the modeled optimum while minimizing the number of expensive samples needed.

## Introduction

Many types of scientific or engineering experiments seek to identify the global optimum (maximum or minimum) of an unknown relationship between a set of experimental inputs $X$ and an output objective $y = f(X)$, where $X$ is a multidimensional vector of input parameters. The "black box" function $f(X)$ is unknown and usually too complicated to be approximated by any specific physics-based parametric representation. In such cases, Bayesian optimization[1] using a data-based non-parametric surrogate regression model has emerged as a powerful and widely adopted machine learning method to guide empirical searches of parameter space seeking the optimal input $X_{opt}$ and the optimal objective value $y_{opt} = f(X_{opt})$. Bayesian optimization is particularly useful when generating new samples to test $f(X)$ is expensive in cost or time, so the campaign success can be achieved with as small a data set as possible, usually a few hundred data points at most. Some examples of Bayesian optimization applications include new materials synthesis and processing,[2–4] mechanical design,[5] new drug discovery,[6] and maximizing manufacturing yield.[7]

In many real experiments, the cost of generating and experimentally evaluating a small batch of $q$ new samples at one time, where usually $q \leq 10$, only marginally exceeds the cost of a single sample. It then makes sense to use batches of $q$ new samples to test $f(X)$ and provide additional data to update the surrogate model in each Bayesian optimization iteration step, a process called batch Bayesian optimization.[8] The goal for real experiments is to maximize confidence in the accuracy of the modeled global optimum using the fewest number of expensive experimental iterations possible.

The key component of batch Bayesian optimization lies in a batch acquisition function that suggests the most promising input parameters to test in the next experimental batch. In each iteration of the process, a chosen acquisition function evaluates existing data, the current surrogate model, and the uncertainty of that model to assess statistically how much new $X$ inputs will contribute towards advancing the search for $y_{opt}$.[9] For non-batch ($q = 1$) cases, the suggested next input is the $X$ that maximizes the acquisition function. Many standard acquisition functions are available for non-batch Bayesian optimization problems, the most common being expected improvement (EI) or its logarithm (logEI)[10] and, for maximization problems, upper confidence bound (UCB).[11]

*[a]Department of Materials Science and Engineering, The University of Texas at Dallas, USA. E-mail: jwhsu@utdallas.edu*

*[b]Department of Physics, The University of Texas at Dallas, USA*

† Electronic supplementary information (ESI) available: ESI Notes 1–3 and Fig. S1–S4. See DOI: https://doi.org/10.1039/d5dd00066a

For batch Bayesian optimization, how to generate a batch of $q > 1$ next $X$ inputs that together most efficiently advance the optimization progress is significantly more challenging.[8] Most batch-picking strategies fall into two general approaches: serial and parallel. Serial batch picking chooses the first $X$ of a batch in the same way as non-batch optimization, then modifies the acquisition function using some strategy to pick a second $X$ that is meaningfully different from the first and iterates the procedure until $q$ new $X$ inputs are assembled. The most common examples of this serial approach include local penalization (LP),[12] and heuristic or "greedy" simplifications of parallel batch acquisition functions known as continuous liar and Kriging believer.[13] Parallel batch picking generalizes a non-batch acquisition function by integrating it over a $q$-point joint probability density function obtained from the surrogate model's covariance kernel.[13–15] The suggested next batch is composed of the $q$ $X$ points that jointly maximize the integrated acquisition function. Examples of $q$-points parallel batch acquisition functions include $q$EI, $q$logEI, and $q$UCB.[15,16]

Serial batch acquisition functions are usually computed and maximized using deterministic numerical methods, *i.e.*, without stochastic sampling. These calculations become computationally more difficult and less accurate when the dimension of $X$ exceeds 5 or 6.[17] Since parallel batch acquisition functions integrate over a probability density function, their calculation and maximization are well suited to be done by stochastic Monte Carlo methods and therefore offer an attractive alternative, especially for high dimensional $X$.[15,17,18] For this reason we call parallel batch acquisition functions such as $q$logEI and $q$UCB "Monte Carlo" acquisition functions.

In this paper, we conduct a direct comparison of serial and parallel batch acquisition functions in batch Bayesian optimization campaigns. The black box functions being optimized include two analytic mathematical functions, one of them evaluated with and without normally distributed noise, that are proxies for input dimensionalities and functional landscapes typically encountered in real experiments on materials synthesis optimization, and one empirical regression model built from real experimental data. The first mathematical model is the Ackley function in 6 dimensions.[19] Ackley epitomizes a "needle-in-haystack" functional landscape because it is a highly heterogeneous function, oscillating near its minimum value through most of its domain except for a sharp peak that occupies a small fraction of its domain hypervolume. The second mathematical model is the Hartmann function, also in 6 dimensions.[20] Hartmann represents a "false maximum" landscape because it has a secondary maximum with an objective value nearly degenerate with its true maximum, but at a different $X$. The empirical model is a 4-dimensional ensemble regression model built using data from an experiment to fabricate flexible halide perovskite solar cells with maximum power conversion efficiency (PCE).[21] This PCE model embodies the real-world difficulties of having only a small number of data points due to the time and cost expense of performing experiments, the inclusion of noise and possible systematic errors that may not be well quantified, and an unknown landscape not guaranteed to be mathematically analytical. Details of this PCE model and its construction are given in ref. 21 and Note 1 in the ESI.† Fig. S1 in the ESI† shows projected maximum ground truth landscapes for all three models.

For the serial batch acquisition function, we use UCB/LP because UCB has been reported to outperform EI or logEI for non-batch Bayesian optimization on a wide range of synthetic functions.[3,22–24] (For completeness, the ESI (Fig. S2†) shows learning performance using log EI/LP). LP is used because it has a sounder intellectual basis than heuristic serial batch picking approaches and, in our experience, outperforms continuous liar and Kriging believer in test cases on synthetic data.[25] For Monte Carlo batch acquisition functions, we use $q$UCB and $q$logEI for noiseless problems, and add a noise-integrated version of $q$logEI called $q$logNEI for evaluations of the Hartmann function with noise as well as for the PCE model since it was built on data with real-world noise.[26] We do not test $q$EI because it offers no advantages over $q$logEI, but is more prone to numerical instability.[27]

$q$UCB is found to give the best overall performance: producing reliable results in all functional landscapes tested, converging with relatively few iterations, and showing reasonable noise immunity. Thus, when the general landscape and noisiness of the black-box function are *a priori* unknown, as is the case for real-world experiments, we recommend $q$UCB as the default acquisition function choice.

## Results and discussion

### Setup of batch Bayesian optimization process

The code used to generate all results shown in this study is publicly available on GitHub (see Data availability). All code was implemented in Python and run in normal mode (CPU only) on the Lonestar6 system of the Texas Advanced Computing Center. We used the Emukit package for UCB/LP (and logEI/LP) and the BoTorch package for the Monte Carlo batch acquisition functions. Computational time and memory usage are given in the ESI (Note 2).†

Fig. 1 depicts a block diagram of the batch Bayesian optimization workflow. In all cases the procedure used a Gaussian process regression surrogate model with an ARD Matern 5/2 kernel. Kernel hyperparameters were optimized by maximizing log-likelihood. For all problems, at each iteration, $X$ training data were normalized to the $[0, 1]^d$ hypercube, where $d = 6$ for Ackley and Hartmann functions and $d = 4$ for the PCE model, and $y$ training data were standardized before computing the posterior surrogate model. New batch selections $X_{new}$ were then unnormalized to make new evaluations, $y = f(X_{new})$, to add to the training data set for the next iteration. For UCB/LP and $q$UCB, the exploration/exploitation parameter $\beta$ was set at 2. Finding the $X$ value that maximizes UCB/LP for each serial pick in the batch was done by a deterministic quasi-Newtonian

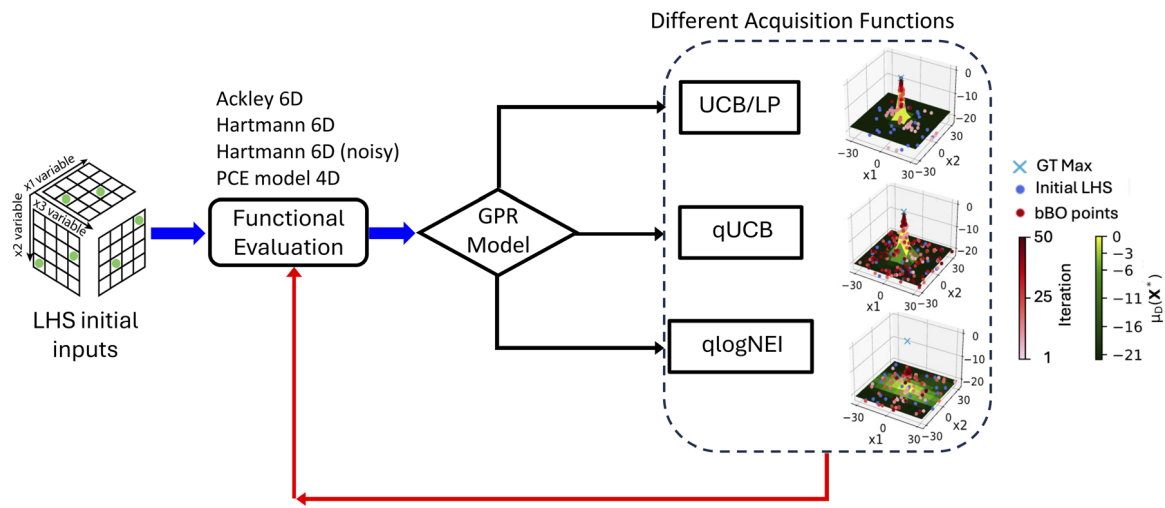

Fig. 1 Block diagram of the Bayesian optimization workflow to compare performance of different batch acquisition functions on various test functions that serve as proxies for functional landscapes typically encountered in real materials synthesis optimization experiments. The initial functional evaluations and Gaussian process regression (GPR) model use Latin Hypercube sampled (LHS) points from each function's domain.

method in Emukit. Finding the $q$ points that jointly maximize the MC batch acquisition functions was done by a stochastic gradient descent method in BoTorch. Reasons for the choices made for these Bayesian optimization settings are given in Note 3 of the ESI.†

In each batch Bayesian optimization campaign, the first surrogate model was built from an initialization training data set $\{X, f(X)\}$ of 24 $X$ points selected from each test function's domain by Latin hypercube sampling. This avoids clustering of $X$ points that can result from purely random sampling and is a commonly used method to select initial processing parameters in materials synthesis experiments when no previous knowledge exists.[28] To collect statistics of learning performance arising from the choice of initial training set, 99 such initialization sets were generated, which were used as the common starting points to test each acquisition function on a given ground truth model. After initialization, in each subsequent Bayesian optimization iteration, the posterior mean surrogate model was updated with additional batches of $q = 4$ data points selected by the batch acquisition function under test and the corresponding surrogate model. The number of iterations in each campaign was capped at 50, so the number of sampled data points is 224 for each run.

For $q$logEI to show learning progression on Ackley, we found it necessary to adaptively narrow the domain search hypervolume in each iteration. Several domain-narrowing methods have been proposed to handle "needle-in-haystack" problems.[29,30] We implemented a trust region BO (TuRBO) strategy coded using the BoTorch package.[31]

The robustness of the batch acquisition functions against output noise was examined with the Hartmann function by adding a normally distributed random value to $f(X)$ in every evaluation of the Hartmann function. The mean of this noise distribution was zero. Noise amplitude was controlled by setting the noise distribution standard deviations to values between 1% to 20% of the Hartmann function's peak-to-peak amplitude.

### Results and discussion on Ackley function

The inverted Ackley function has its true maximum at $X_{\max} = [0, 0, 0, 0, 0, 0]$ with a true maximum objective value $y_{\max} = f(X_{\max}) = 0$. Its domain is the hypercube $[-32.768, 32.768]^6$, with side length $L = 65.536$, and its range is $[-22.3, 0]$, giving peak-to-peak amplitude $\Delta y = 22.3$. A maximum projection surface plot of the Ackley function is shown in Fig. S1(a) in the ESI.† Ackley is a highly heterogeneous function; $y < -18$ through the vast majority of its domain, with a large central peak centered on $y_{\max}$ that occupies only ∼0.08% of its domain hypervolume and drops steeply from its central maximum.

Fig. 2 summarizes the batch Bayesian optimization learning progression on Ackley for UCB/LP compared to $q$UCB, $q$logEI, and $q$logEI + TuRBO. Each plot shows results from all runs for each acquisition function under test starting from the same 99 initial training data sets. In Fig. 2 $\mu(X^*)$ is the maximum objective value predicted by the mean posterior surrogate model $\mu(X)$, and $X^*$ is the input vector that produces the maximum $\mu(X)$, up through the $n$th iteration. The left column plots Fig. 2(a)–(d) show $\mu(X^*)$ at each iteration relative to $y_{\max} = 0$, indicated by the yellow dashed line. Fig. 2(e)–(h) show the Euclidean distance magnitude between $X^*$ and $X_{\max}$ at each iteration, so zero indicates the model has found $X_{\max}$. After the final (50th) iteration, the 99 runs are percentile ranked by how close the final $\|X^* - X_{\max}\|$ is to zero, with the 99th percentile being the best. Green, red, and blue points highlight the runs ranked 25th, 50th, and 75th percentile, respectively.

From Fig. 2, it is clear that UCB/LP and $q$UCB perform comparably well, with both significantly outperforming $q$logEI. In terms of finding $X_{\max}$, Fig. 2(e) and (f) show that $\|X^* - X_{\max}\|$ for UCB/LP and $q$UCB both converge to near zero, within a few percent of the domain hypercube side length $L = 65.536$, in fewer than ∼15 iterations for nearly all 99 initial sets. In terms of how accurately $y_{\max}$ can be modeled, Ackley presents a difficult challenge for surrogate models because its maximum sits on a very steep peak, so relatively small values of $\|X^* - X_{\max}\|$ can yield $\mu(X^*) \ll y_{max}$, causing the surrogate model to systematically underestimate $y_{max}$. In Fig. 1(a), UCB/LP shows the best final $\mu(X^*)$ estimate of $y_{max}$, converging to within ∼2 of $y_{max}$ (<10% of the amplitude $\Delta y = 22.3$) for all 99 initial conditions by the final iteration. $q$UCB performs a close second, converging to within ∼3 of $y_{max}$ (<15% of $\Delta y$) for all 99 initial conditions.

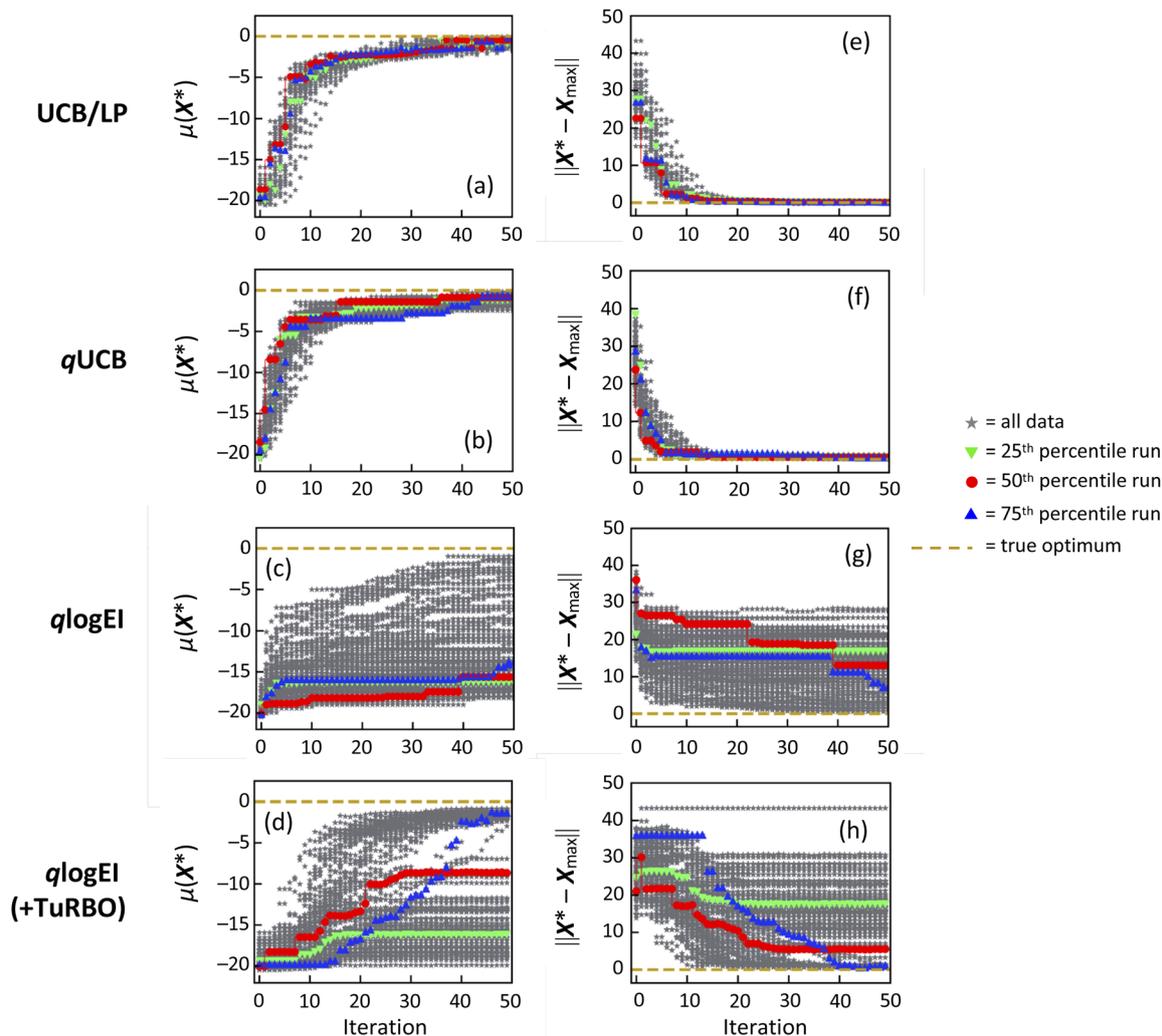

**Fig. 2** Learning progression data on Ackley comparing UCB/LP against $q$UCB, $q$logEI, and $q$logEI + TuRBO. Plots (a)–(d) are $\mu(X^*)$, the maximum values of the mean posterior surrogate model, with $X^*$ being the input value that produces the maximum $\mu(X)$, up to that iteration of the campaign. Plots (e)–(h) are Euclidean distance magnitudes between the true $X_{max}$ and $X^*$. The ground truth $X_{max}$ and $y_{max}$ are indicated by yellow dashed lines. Gray points show the spread in learning progress of the 99 batch Bayesian optimization runs starting from the 99 LHS initial data sets. Green, red, and blue points indicate the runs ranked in the top 25th, 50th, and 75th percentile, respectively.

By contrast, Fig. 2(c) and (g) show that $q$logEI fails to model anything close to $y_{max}$ or $X_{max}$ after 50 iterations for most of the 99 initial conditions. In fact, $q$logEI essentially fails to show further learning progress after 5 to 10 iterations in most cases. This is consistent with ref. 24, where $q$EI failed to model 6-dimensional Ackley. Upon augmenting with TuRBO, Fig. 2(d) and (h) show that learning is partially restored for $q$logEI for many, but far from all, 99 initial conditions. While $q$logEI with TuRBO can converge within roughly 3 of $y_{max}$, there is much larger variation depending on the initial conditions. The performance of UCB/LP and $q$UCB remain obviously superior to $q$logEI even with TuRBO.

Table 1 summarizes the learning progression graphically depicted in Fig. 2 using the quantitative metrics: instantaneous regret (IR), which measures how accurately the final optimal point is modeled, and cumulative regret (CR), which measures how fast the batch Bayesian optimization process converges onto the optimal point, in both $y$ and $X$. The values given in Table 1 are averaged over all runs from the 99 initial conditions. IR($y$) and CR($y$) are normalized to the Ackley amplitude, $\Delta y = 22.3$, and IR($X$) and CR($X$) are normalized to the Ackley domain side $L = 65.536$. The closer IR and CR are to zero, the better the final accuracy and convergence rate of the process.

The metrics in Table 1 show that while UCB/LP generates the best average final surrogate model, both UCB/LP and $q$UCB

Table 1 Summary of normalized instantaneous regret (IR) and normalized cumulative regret (CR) in $y$ and $X$ on Ackley for each batch acquisition function, averaged over the results of all 99 campaigns starting with different initial data sets. Box and violin plots visualizing the IR and CR distributions for the 99 campaigns in each case are given in the ESI (Fig. S3a)

| Acq. Fn. | $\langle IR(y)\rangle/\Delta y$ | $\langle CR(y)\rangle/\Delta y$ | $\langle IR(X)\rangle/L$ | $CR(X)/L$ |
| --- | --- | --- | --- | --- |
| UCB/LP | 0.017 | 4.9 | 0.0016 | 1.2 |
| $q$UCB | 0.026 | 5.1 | 0.016 | 2.7 |
| $q$logEI | 0.56 | 32 | 0.18 | 11 |
| $q$logEI + TuRBO | 0.34 | 17 | 0.14 | 6.5 |

produce accurate and reliable estimations of the true Ackley maximum within 20 to 30 iterations, independent of initial conditions. For both, $\langle IR(X)\rangle/L$ and $\langle IR(y)\rangle/\Delta y$ are $\ll 1$.

An interesting question is why UCB/LP outperforms $q$UCB on Ackley, especially in producing a nearly perfect $\langle IR(X)\rangle/L$ metric. A possible answer lies in the stochastic nature of Monte Carlo based compared to deterministic serial batch-picking algorithms. Local penalization (LP) adaptively becomes more exploitative and less explorative as new data become available, biasing its batch picks to increase sampling density in domain regions that generate higher objective values.[12] Consequently, as soon as one point in the steep central maximum region is found, LP biases all subsequent batch picks to exploit that domain region in greater detail, giving a better surrogate model reconstruction of Ackley's central peak and hence better performance as measure by IR($y$) and IR($X$). By contrast, the stochastic nature of Monte Carlo evaluation and optimization of $q$UCB results in greater scattering of batch picks in all iterations. Even after one point in the central maximum region is found, $q$UCB may assign only a single new point in the next iteration batch to exploit the nearby region and stochastically scatter other batch points to explore the domain. As a result, the region near the maximum is not tested in as much detail compared to UCB/LP, giving a less accurate surrogate model reconstruction of the central peak, though possibly a better model of the overall function. Fig. 3 shows a graphic example of this difference in sampling distribution between UCB/LP and $q$UCB, and a time series showing batch picks and surrogate model after each iteration is shown in Fig. S4 in the ESI.†

### Results and discussion on Hartmann function without noise

The inverted Hartmann test function has its true maximum at $X_{\max} = [0.2017, 0.1500, 0.4769, 0.2753, 0.3117, 0.6573]$ with a maximum objective value $y_{\max} = f(X_{\max}) = 3.3224$. Its domain is the hypercube $[0, 1]^6$, with side length $L = 1$, and its range is $[0, 3.3224]$, giving peak-to-peak amplitude $\Delta y = 3.3224$. A maximum projection surface plot of the Hartmann function is shown in Fig. S1(b) in the ESI.† Hartmann is complicated by the existence of a secondary maximum at $X_2 = [0.4047, 0.8824, 0.8461, 0.5740, 0.1390, 0.0385]$ whose objective value $y_2 = f(X_2) = 3.2032$ is nearly degenerate with $y_{\max}$; the distance $\|X_2 - X_{\max}\| = 1.10$. Consequently, maximization searches can easily converge onto the "false maximum" $X_2$ rather than $X_{\max}$.

Fig. 4 summarizes the learning progression on Hartmann for UCB/LP compared to $q$UCB and $q$logEI. The use of TuRBO was unnecessary because $q$logEI works for Hartmann. Each plot shows the results of all batch Bayesian optimization runs starting from the common set of 99 initial data sets in the Hartmann domain. The meaning of all terms and symbols is the same as for Fig. 2. The left column plots (a)–(c) show $\mu(X^*)$ at each iteration relative to $y_{\max} = 3.3224$, which is indicated by the yellow dashed line. Fig. 4(d)–(f) show the Euclidean distance magnitude between $X^*$ and $X_{\max}$ at each iteration. One obvious feature of Fig. 4(d)–(f) is that $X$ results for all three acquisition functions bifurcate, converging upon two different best $X$ points. This is a consequence of the existence of a second maximum with a nearly degenerate objective value in Hartmann. Some initial conditions lead to convergence onto $X_2$. This result is not as obvious in Fig. 4(a)–(c) because $y_2$ is only 0.12 less than $y_{max}$. Because of the false maximum nature of the Hartmann landscape, an additional batch acquisition function performance metric is the percentage of the 99 initial conditions that converge onto $X_2$ instead of $X_{max}$ in Fig. 4(d)–(f), with a smaller percentage being better, shown in the rightmost column of Table 2.

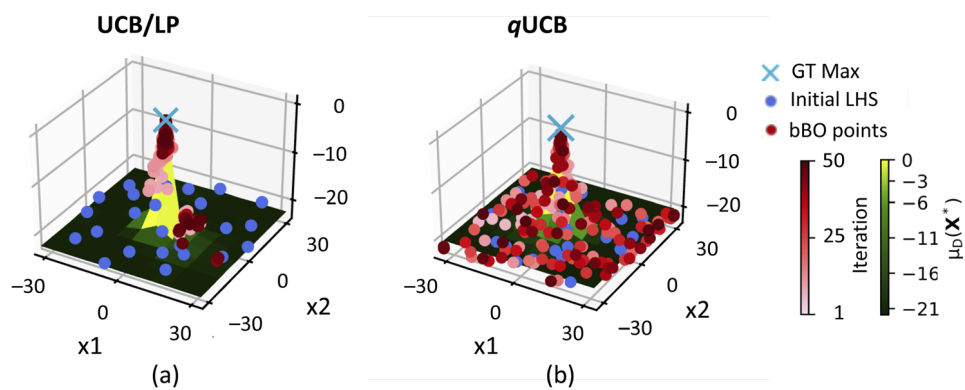

Fig. 3 Distribution of all sampled points picked by (a) UCB/LP and (b) $q$UCB in the $x_1$, $x_2$ plane, where the input vector $X = (x_1, x_2, x_3, x_4, x_5, x_6)$. The underlying $z$-axis contour plot shows the shape of the final surrogate model $\mu(X)$ for the initial data set that achieves the 50th percentile outcome (red curves in Fig. 1) by projecting the maximum value of $\mu(X)$ evaluated at each $x_1$, $x_2$ coordinate. The blue "×" indicates the ground truth maximum (GT Max) of the Ackley function. Input points belonging to the initial sampling (initial LHS) are shown as blue circles. Batch Bayesian optimization points (bBO points) picked by the batch acquisition functions are shown as reddish circles, with light red/pink indicating points picked in early iterations and darker red circles indicating points picked in later iterations.

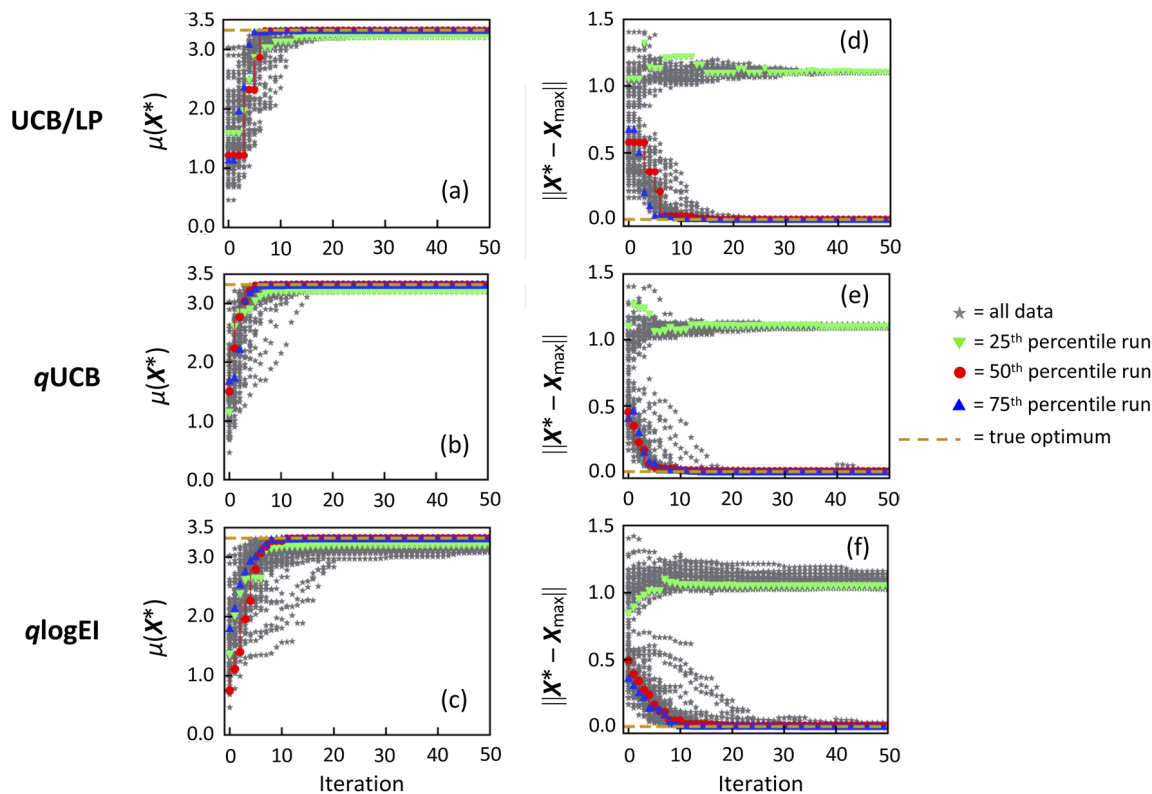

**Fig. 4** Learning progression data on the noiseless Hartmann test function comparing UCB/LP against $q$UCB and $q$logEI. All variable labels and symbols have the same meanings as in Fig. 2.

Fig. 4 shows that all three batch acquisition functions converge to a final value of $\mu(X^*)$ near $y_{max}$ and a final $X^*$ near either $X_{max}$ or $X_2$ within ∼20 iterations. Visually, $q$UCB appears to converge the fastest, in <10 iterations for most initial conditions. After ∼15 iterations both UCB/LP and $q$UCB become insensitive to initial conditions, except for the bifurcation. $q$logEI appears to converge more slowly and clearly has a larger performance spread depending on initial conditions compared to UCB/LP and $q$UCB.

**Table 2** Summary of normalized instantaneous regret (IR) and normalized cumulative regret (CR) in $y$ and in $X$ on Hartmann for each batch acquisition function, averaged over the results of all 99 campaigns starting with different initial data sets. Box and violin plots visualizing the IR and CR distributions for the 99 campaigns in each case are given in the ESI (Fig. S3b)

| Acq. Fn | $\langle IR(y)\rangle/\Delta y$ | $\langle CR(y)\rangle/\Delta y$ | $\langle IR(X)\rangle/L$ | $\langle CR(X)\rangle/L$ | % False max |
|---|---|---|---|---|---|
| UCB/LP | 0.0081 | 3.3 | 0.24 | 19 | 30 |
| $q$UCB | 0.012 | 2.1 | 0.36 | 20 | 32 |
| $q$logEI | 0.015 | 3.1 | 0.37 | 20 | 34 |

Table 2 summarizes the learning performance metrics on Hartmann: normalized IR($y$), CR($y$), IR($X$), and CR($X$), averaged over all runs from the 99 LHS initial conditions, and the percentage of the LHS initial conditions that converge onto the false maximum. IR($y$) and CR($y$) are normalized to the Hartmann range amplitude, $\Delta y = 3.3224$. IR($X$) and CR($X$) do not technically need to be normalized because the domain hypercube side $L = 1$, but are listed as normalized to $L$ for consistency with Table 1. In all columns, smaller numerical values indicate better final accuracy, convergence rate, and convergence onto the true maximum of the batch Bayesian optimization process.

The $\langle IR(y)\rangle/\Delta y$ and $\langle CR(y)\rangle/\Delta y$ values in Table 2 are generally smaller compared to the same metrics for Ackley (see Table 1), but $\langle IR(X)\rangle/L$ and $\langle CR(X)\rangle/L$ are significantly larger for all acquisition functions relative to Ackley. This again reflects the false maximum nature of Hartmann. For each acquisition function, roughly one-third of the initial conditions produce Bayesian optimization campaigns that converge onto the false maximum. Each of these runs contributes a regret of $\|X_2 - X_{max}\| = 1.10$ towards the IR($X$)/$L$ and CR($X$)/$L$ averages but contributes a regret of only 0.036 towards the IR($y$)/$\Delta y$ and CR($y$)/$\Delta y$ averages.

The metrics in Table 2 show that UCB/LP and $q$UCB perform very similarly on Hartmann without noise, with UCB/LP doing slightly better in $\langle IR(y)\rangle/\Delta y$, $\langle IR(X)\rangle/L$, and $\langle CR(X)\rangle/L$ while $q$UCB shows somewhat better $\langle CR(y)\rangle/\Delta y$. Although $q$logEI shows reasonable learning behavior unlike for Ackley, both UCB/LP and $q$UCB outperform $q$logEI on noiseless Hartmann in all metrics.

### Results and discussion on Hartmann function with noise

Learning progression plots on Hartmann with noisy functional evaluations are shown in Fig. 5 for 5% noise amplitude and Fig. 6 for 20% noise amplitude. In addition to the three batch acquisition functions investigated for the noiseless Hartmann study in Fig. 4, included here is $q$logNEI, which is $q$logEI integrated over a normally distributed noise probability, which is designed to deal specifically with noise.[26] Visually comparing the plots in Fig. 5, at a moderate 5% noise, all acquisition functions still model a final value of $\mu(X^*)$ close to $y_{max}$ and converge close to $X_{max}$ or $X_2$ for most initial conditions, although there is degradation in learning performance compared to the noiseless case (Fig. 4). From Fig. 6, at very high 20% noise, all Monte Carlo acquisition functions clearly outperform UCB/LP. A far higher number of initial conditions converge to something reasonably close to true maximal values in both $y$ and $X$ for the three Monte Carlo acquisition functions compared to UCB/LP.

Fig. 7 summarizes the dependence of these learning results on noise amplitude for UCB/LP, $q$UCB, $q$logEI, and $q$logNEI for noisy Hartmann. Fig. 7(a), (b) plot regrets in $y$ and Fig. 7(c), (d) plot regrets in $X$, all vs. noise amplitude. These measure degradation in how well $y_{max}$ and best $X_{max}$ are modeled and in convergence onto the optimal values as noise increases.

At 5% noise amplitude, Fig. 5 and 7 show that after 50 iterations, all acquisition functions converge reasonably close to $y_{max}$, with UCB/LP and $q$UCB slightly outperforming both $q$logEI and $q$logNEI in $\langle IR(y) \rangle / \Delta y$ and $\langle CR(y) \rangle / \Delta y$. Also, all acquisition functions converge on $X_{max}$ or $X_2$, with UCB/LP giving lower $\langle IR(X) \rangle / L$ and $\langle CR(X) \rangle / L$ than the Monte Carlo acquisition functions. However, looking at the $X$ learning plots Fig. 5(e)–(h), UCB/LP shows greater sensitivity to initial conditions than the Monte Carlo acquisition functions. UCB/LP's lower $\langle IR(X) \rangle/L$ and $\langle CR(X) \rangle/L$ values at 5% noise mostly stem from the fact that, for those runs converging onto $X_2$, UCB/LP converges to an $\|X_2 - X_{max}\|$ value < the true 1.10 on average, while the Monte Carlo functions converge to an $\|X_2 - X_{max}\|$ >1.10 on average. This is evident from the values of the upper ($X_2$) branch in Fig. 5(e)–(h). For those runs converging onto $X_{max}$, by the 50th iteration the performance of all acquisition functions are nearly equal.

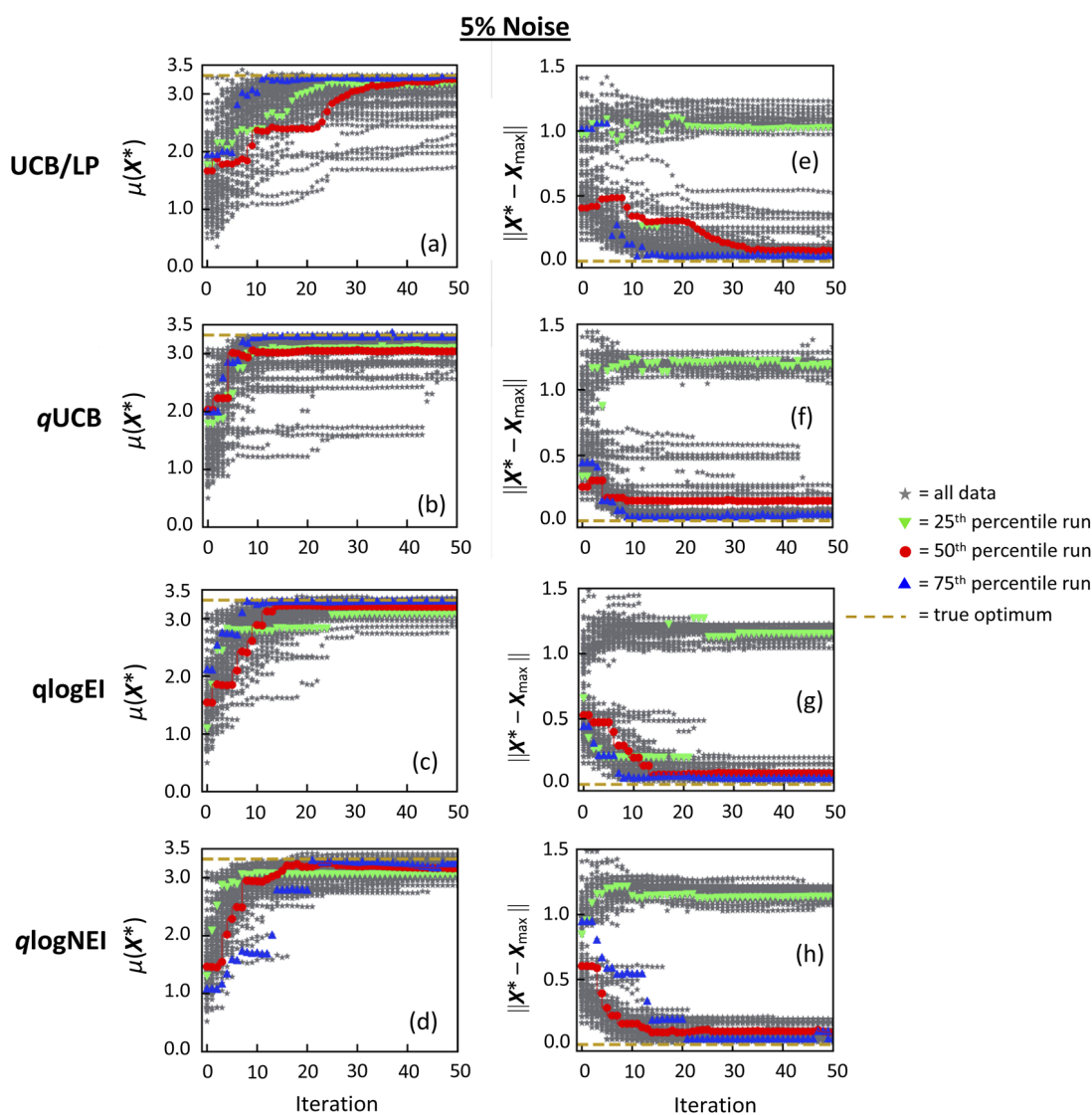

Fig. 5 Learning progression data on Hartmann with noisy functional evaluations at 5% noise amplitude. All variable labels and symbols have the same meaning as in Fig. 2.

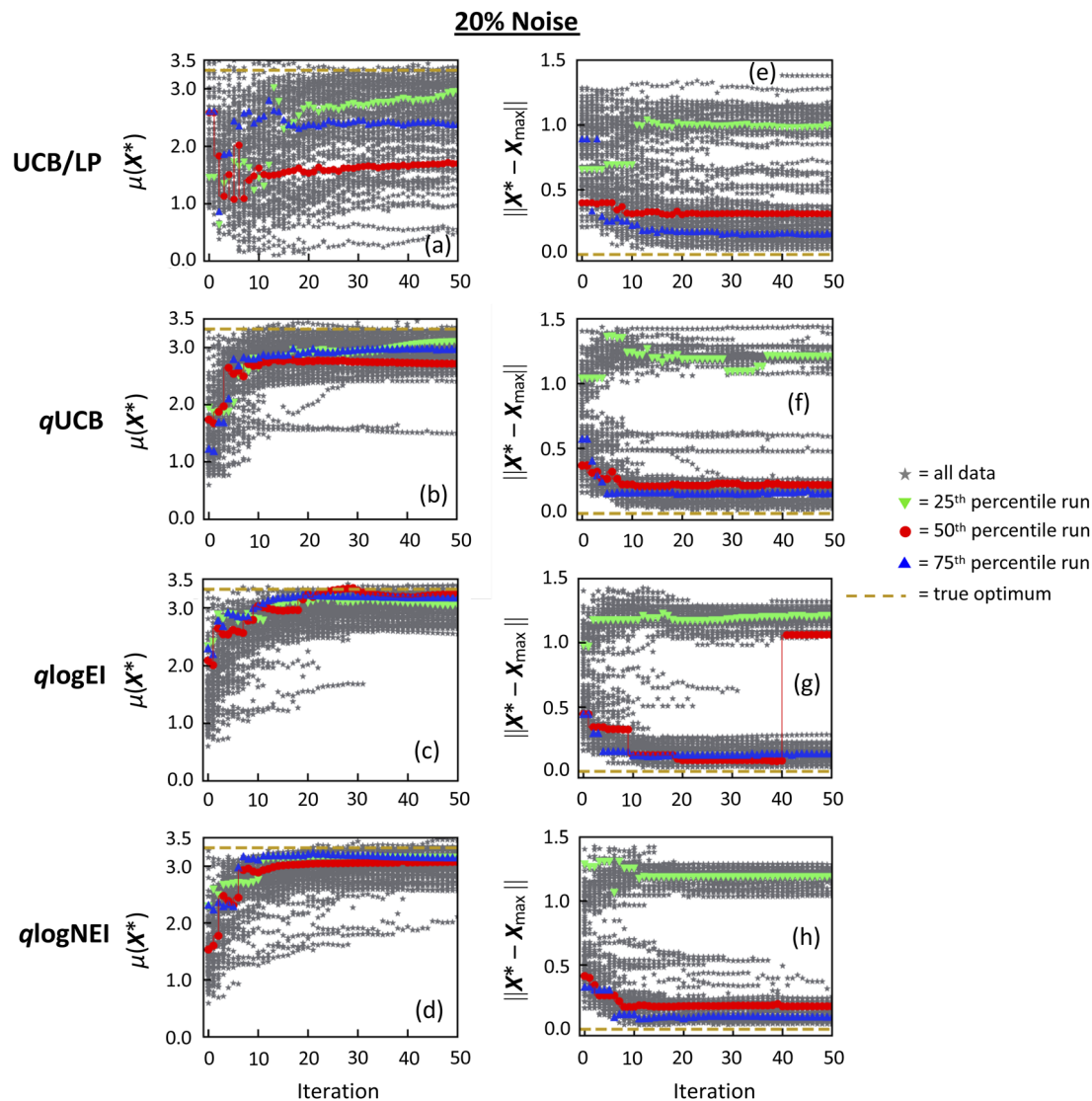

**Fig. 6** Learning progression data on Hartmann with noisy functional evaluations at 20% noise amplitude. All variable labels and symbols have the same meaning as in Fig. 2.

At high noise amplitude of 20%, Fig. 6 and 7 show that all Monte Carlo acquisition functions perform better than UCB/LP, although with clearly degraded accuracy. The three Monte Carlo acquisition functions do a reasonable job of modeling the value of $y_{max}$ and finding $X_{max}$ or $X_2$ for the large majority of initial conditions. By contrast, Fig. 6(a) shows that at 20% noise level, UCB/LP nearly fails to model $y_{max}$, and Fig. 6(e) shows that UCB/

LP is much more sensitive to initial conditions in modeling $X_{max}$ compared to the Monte Carlo functions. Fig. 7(a) and (b) reflect the fact that at higher noise levels UCB/LP is significantly worse at modeling $y_{max}$ accurately and in convergence onto an optimal objective. At high noise, $q$UCB shows the best performance overall in the $y$-regret metrics. Fig. 7(c) and (d) show UCB/LP appears to have slightly better $X$ regret metrics even up to 20% noise, but again this is mostly a result of UCB/LP systematically underestimating $\|X_2 - X_{max}\|$ while the Monte Carlo functions tend to overestimate $\|X_2 - X_{max}\|$.

From Fig. 7, all Monte Carlo acquisition functions behave similarly with regard to noise level on Hartmann, with $q$UCB showing a slight advantage. Perhaps the stochastic nature of the Monte Carlo computations partially compensates for the randomness of noisy functional evaluations. It should be noted that $q$logNEI was developed specifically for noisy batch Bayesian optimization problems to increase surrogate model accuracy and decrease model sensitivity to noise, but in these tests, $q$logNEI showed no significant advantage over $q$UCB.

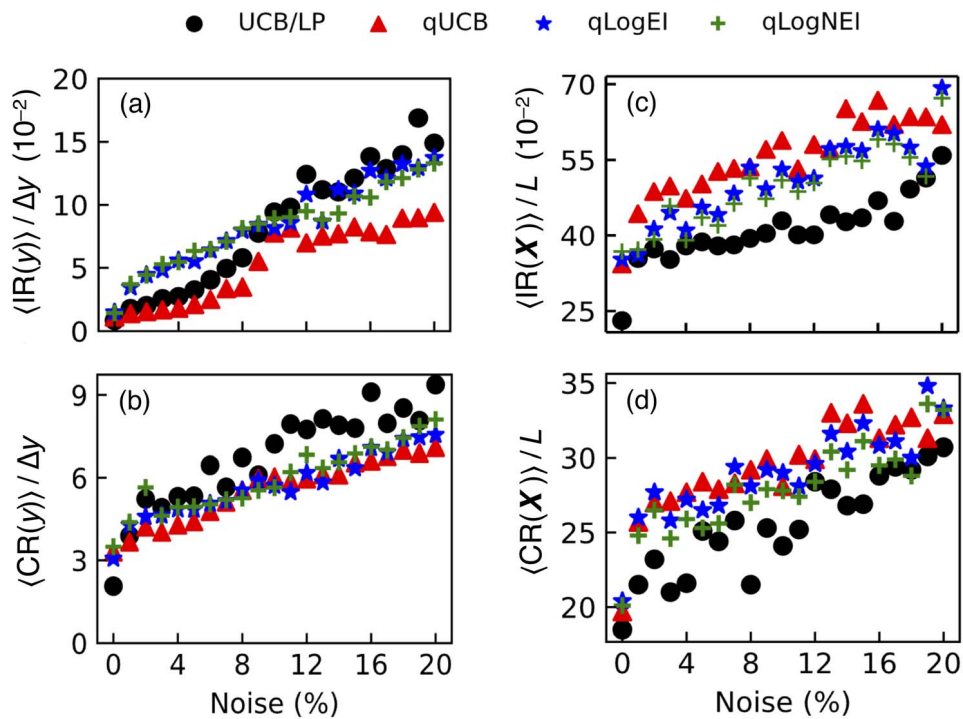

Fig. 7 Performance of batch acquisition function (UCB/LP, qUCB, qlogEI, and qlogNEI) against noise magnitude in batch evaluations of Hartmann. Left column: optimality and convergence in modeled best objective value $y$ as measured by (a) instantaneous regret and (b) cumulative regret. Right column: optimality and convergence in modeled best predictor value $X$ as measured by (c) instantaneous regret and (d) cumulative regret. All regrets are averaged and normalized as described in the text.

### Results on empirical perovskite PCE model

To evaluate how these batch acquisition functions would perform on a real experimental optimization problem as opposed to analytic mathematical functions, we used a fully trained non-parametric regression model built from experimental data on maximizing power conversion efficiency (PCE) of flexible perovskite solar cells as the ground truth function. Details of the experiment and construction of this "PCE model" are given in ref. 21 and Note 1 in the ESI.† The PCE model is 4-dimensional with true $y_{max} = 11.2$ at $X_{max} = [0.40, 0.60, 0.40, 0.21]$ in the normalized domain hypercube of $[0, 1]^4$, so the domain length $L = 1$. Its $y$ range is $[0.5, 11.2]$, giving peak-to-peak amplitude $\Delta y = 10.7$. Truth function values from the PCE model were evaluated on a gridded domain with grid spacing in each dimension of $X$ determined by the experimental step size used for each input predictor. Truth function values are stored as a look-up table, available on-line (see Data availability). In the batch Bayesian optimization process, functional evaluations for each batch of recommended next $X$ values were drawn from this look-up table.

Fig. 8 summarizes learning results on the PCE model for UCB/LP, $q$UCB and $q$logNEI, which was used because the PCE model was built using data with real noise and it is of interest to see whether $q$logNEI offers any advantages in a real-world noise scenario. Learning progression data for $\mu(X^*)$ and $\|X^* - X_{max}\|$ for each batch acquisition function under test are shown in Fig. 8(a)–(f). Each of these plots shows the results of all batch Bayesian optimization runs starting from the common set of 99 initial data sets in the PCE model domain. The meaning of all terms and symbols is the same as for Fig. 2. The left column plots Fig. 8(a)–(c) show $\mu(X^*)$ for each acquisition function at each iteration relative to $y_{max} = 11.2$, which is indicated by the yellow dashed line. The middle column plots Fig. 8(d)–(f) show the Euclidean distance magnitude between $X^*$ and $X_{max}$ for each batch acquisition function at each iteration. Most notably, the convergence of $X^*$ to $X_{max}$ for all three acquisition functions is non-zero and strongly dependent on initial conditions, similar to the synthetic functions with noise (Fig. 5 and 6).

The reason for this apparent poor learning performance in $\|X^* - X_{max}\|$ is shown in Fig. 8(g)–(i). These are heat map plots of the ground truth landscape made by projecting the maximum value of the ground truth function at each $X = (x_1, x_2, x_3, x_4)$ onto the $x_1, x_2$ plane. Surface plots of this landscape are shown in the ESI (Fig. S1(c)).† This landscape has a broad, nearly flat plateau near the maximum PCE in the $x_1, x_2$ plane. Consequently, the batch Bayesian optimization process can stop at many different $X^*$ inputs that give $\mu(X^*)$ values very close to $y_{max}$. Comparing Fig. 8(g)–(i) shows that the $X^*$ from the model (color circles) are closest to $X_{max}$ with $q$UCB as the batch acquisition function. Additionally, real-world experimental results inherently contain non-Gaussian and unknown systematic errors and uncertainties.

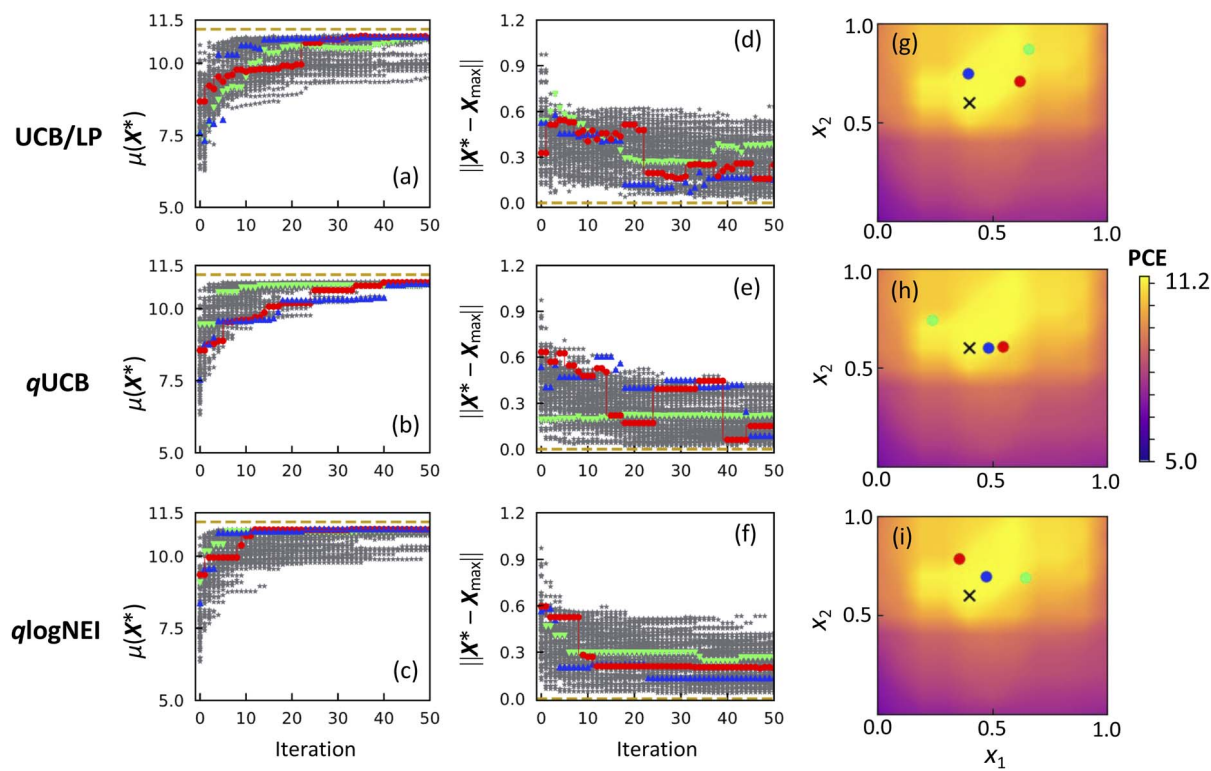

**Fig. 8** Summary of learning results on the PCE model comparing UCB/LP (top row) against $q$UCB (middle row) and $q$logNEI (bottom row). For learning progression plots (a)–(f), all variable labels and symbols have the same meaning as in Fig. 2. Plots (g)–(i) are heat map plots of the ground truth function made by projecting the maximum value the ground truth function at each $X = (x_1, x_2, x_3, x_4)$ onto the $x_1$, $x_2$ plane. The true maximum is marked by an "×". The $x_1$, $x_2$ coordinates of the final $X^*$ points found by each batch acquisition function are indicated by green (25th percentile run), red (50th percentile run), and blue (75th percentile run) dots. A surface contour plot of the PCE model landscape is shown in Fig. S1(c) in the ESI.†

In this experiment, getting a PCE value close to $y_{max}$ (i.e., within experimental uncertainty or reproducibility) regardless of $X$ is more important than finding $X_{max}$. For this reason, the $\mu(X^*)$ learning plots should outweigh the $\|X^* - X_{max}\|$ learning plots. From Fig. 8(a)–(c), $q$UCB shows the best overall performance, finding PCE values within ~10% of $y_{max}$ in ~30 iterations and within ~2% of $y_{max}$ in ~40 iterations for all 99 LHS initial conditions. UCB/LP and $q$logNEI also find PCE values within ~2% of $y_{max}$ in ~50 iterations but converge more slowly and do significantly worse on a significant fraction of the 99 initial conditions compared to $q$UCB.

These observations are quantified by the IR and CR metrics shown in Table 3. All metrics are normalized and averaged over the 99 runs for each batch acquisition function tested. As expected, all $\langle IR(X)\rangle/L$ and $\langle CR(X)\rangle/L$ values are large due to the broad flat near-maximum plateau of the landscape. Consistent with the above discussion, $q$UCB shows better performance in finding near-optimum objective values as measured by $\langle IR(y)\rangle/\Delta y$ and converges onto a near-optimum objective value somewhat faster than $q$logNEI and significantly faster than UCB/LP as measured by $\langle CR(y)\rangle/\Delta y$ values.

## Conclusions

Batch Bayesian optimization is a useful machine learning tool to guide real-world scientific and engineering experiments towards cost-effectively searching input parameter space $X$ to find the optimal objective value for an unknown black-box functional relationship $y = f(X)$. The goal of a batch Bayesian optimization campaign is to produce a data-based regression model that can model $y_{opt}$ and $X_{opt}$ with high confidence while using the minimum number of expensive evaluations of $f(X)$ possible.

Critical to achieving this goal is choosing a batch acquisition function. Unfortunately, literature provides little advice on what the "best" choice might be at the start of a Bayesian optimization campaign when little or nothing is known about the black box function being optimized. To provide some empirical guidance, this paper presents results of a direct comparison between a widely used and effective serial batch acquisition function, UCB/LP, computed by standard deterministic numerical methods, against a set of Monte Carlo based parallel acquisition functions, $q$UCB, $q$logEI, and $q$logNEI (for test cases with noise).

**Table 3** Summary of normalized instantaneous regret (IR) and normalized cumulative regret (CR) in $y$ and in $X$ on the empirical PCE model for each acquisition function, averaged over the results of all 99 campaigns starting with different initial data sets. Box and violin plots visualizing the IR and CR distributions for the 99 campaigns in each case are given in the ESI (Fig. S3c)

| Acq. Fn. | $\langle IR(y)\rangle/\Delta y$ | $\langle CR(y)\rangle/\Delta y$ | $\langle IR(X)\rangle/L$ | $\langle CR(X)\rangle/L$ |
|---|---|---|---|---|
| UCB/LP | 0.038 | 4.2 | 0.26 | 17 |
| $q$UCB | 0.026 | 2.7 | 0.18 | 13 |
| $q$logNEI | 0.030 | 2.8 | 0.21 | 14 |

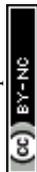

The test problems used, Ackley, Hartmann, and the PCE Model, are proxies for real materials synthesis and optimization experiments in terms of the number of input dimensions, functional landscapes, and noise levels. UCB/LP and $q$UCB do very well on Ackley, with UCB/LP overall performing slightly better than $q$UCB. $q$logEI, on the other hand, struggles to correctly model Ackley even with assistance by an adaptive domain-narrowing algorithm. On noiseless Hartmann, all acquisition functions perform adequately. UCB/LP and $q$UCB perform similarly to each other while both outperform $q$logEI. On Hartmann with noise, all Monte Carlo acquisition functions outperform UCB/LP, particularly at very high noise levels where $q$UCB shows better ability to accurately model the objective maximum. Although $q$logNEI was developed specifically to handle noisy functional evaluations, it shows no clear performance advantage over $q$UCB. On the PCE model, $q$UCB finds input conditions giving near-optimum PCE values in fewer iterations and is less sensitive to initial conditions compared to UCB/LP and $q$logNEI.

In the real-world materials optimization experiments this work is meant to emulate, usually nothing is known *a priori* about the general landscape of the functional relationship $f(X)$ being tested, and empirical evaluations of $f(X)$ always include noise, though the noise level and its probability distribution may not be known. Our results suggest that for batch Bayesian optimization, $q$UCB overall outperforms its Monte Carlo cousins $q$logEI and $q$logNEI as well as its serial version UCB/LP on needle-in-haystack, false optima, and a real experimental functional landscape, and against moderate to high levels of normally distributed noise and unquantified real-world noise. We note that ref. 24 settled on $q$UCB as the batch acquisition function best suited to Bayesian optimization applied to computational fluid dynamics problems, although they did not try UCB/LP, considered only relatively small Gaussian noise amplitudes, and did not test on models built from real data. In our work, $q$UCB is recommended as the default choice of batch acquisition function when applying Bayesian optimization to materials synthesis experiments, at least up to 6 input dimensions, when minimizing the number of expensive iterations and maximizing confidence in the correctness of the result are important.

## Data availability

All codes used to generate the data plotted in this paper and the truth table for the PCE model are available on Github at URL: github.com/UTD-Hsu-Lab/BoTorch-vs-LP (https://github.com/UTD-Hsu-Lab/BoTorch-vs-LP).

## Author contributions

I. M.: data curation, formal analysis, investigation, software, validation, visualization. M. L.: conceptualization, methodology, validation, writing – original draft preparation, writing – review & editing. W. X.: data curation, validation, visualization, writing – review & editing. W. V.: software, supervision, writing – review & editing. J. W. P. H.: funding acquisition, project administration, resources, supervision, validation, writing – review & editing.

## Conflicts of interest

There are no conflicts to declare.

## Acknowledgements

This work is supported by the National Science Foundation CMMI-2135203. We acknowledge the Texas Advanced Computing Center (TACC) at the University of Texas at Austin for providing the high-performance computing resources that have contributed to the research results reported within this paper. URL: https://www.tacc.utexas.edu. J. W. P. H. acknowledges the support of the Texas Instruments Distinguished Chair in Nanoelectronics.